\documentclass[12pt,a4paper]{article}
\usepackage{graphicx}
\usepackage[cp1251]{inputenc}
\usepackage{amssymb, amsfonts, amsmath}
\numberwithin{equation}{section}
\date{}
\usepackage{amscd}
\begin{document}

\title{Maxwell's equations in homogeneous spaces for admissible electromagnetic fields}

 \author{Valery V. Obukhov}

 \maketitle

 \quad

Institute of Scientific Research and Development,
Tomsk State Pedagogical University. Tomsk State  Pedagogical University, 634041, Tomsk, Russia;

  Laboratory for Theoretical Cosmology, International Center of Gravity and Cosmos, Tomsk State University of Control Systems and Radio Electronics, 634050, Tomsk,  Russia.

 \quad


 \quad

\section{Introduction} A special place in mathematical physics is occupied by the problem of exact integration of the equations of motion of a classical or quantum test particle in external electromagnetic and gravitational fields. This problem is closely related to the study of the symmetry of gravitational and electromagnetic fields in which a given particle moves. A necessary condition for the existence of such symmetry is the admissibility of the algebra of symmetry operators, given by vector and tensor Killing fields, for spacetime and the external electromagnetic field. The algebras of these operators are isomorphic to the algebras of the symmetry operators of the equations of motion of a test particle - Hamilton-Jacobi, Klein-Gordon-Fock, or Dirac-Fock. At present, two methods are known for the exact integration of the equations of motion of a test particle. These are the methods of commutative and noncommutative integration.
The first method is based on the use of a commutative algebra of symmetry operators (integrals of motion) that form a complete set.
The complete set includes linear operators of first and second degree in momentum formed by vector and tensor Killing fields of complete sets of geometric objects of \quad $V_4$. \quad
The method is known as the method of complete separation of variables (in the Hamilton-Jacobi, Klein-Gordon-Fock, or Dirac-Fock equations). The spaces in which the method of complete separation of variables is applicable are called Stackel spaces. The theory of Stackel spaces was developed in \cite{1}-\cite{8}. A description of the theory and a detailed bibliography can be found in \cite{9}], \cite{10}-\cite{12}. The most frequently used exact solutions of the gravitational field equations in the theory of gravity were constructed on the basis of Stackel spaces (see, e.g., \cite{13}-\cite{14}). These solutions are still widely used in the study of various effects in gravitational fields (see, e.g., \cite{17}- \cite{19} ).

The second method (noncommutative integration) was developed in \cite{20}. This method is based on the use of an algebra of symmetry operators, which are linear in momenta and constructed using Killing vector fields forming noncommutative groups of motion of spacetime \quad $G_3$ \quad and \quad $G_4$.\quad
The algebras of the symmetry operators of the Klein-Gordon-Fock equation constructed using the algebras of the operators of the noncommutative motion group of space  \quad $V_4$, \quad are complemented to a commutative algebra by the operators of differentiation of the first order in \quad $4$ \quad essential parameters. Among these spacetime manifolds, the homogeneous spaces are of greatest interest for the theory of gravity (see, e.g., \cite{21}-\cite{25}).

Thus, these two methods complement each other to a considerable extent and have similar classification problems (by solving the classification problem, we mean enumerating all metrics and electromagnetic potentials that are not equivalent in terms of admissible transformations).
Among these classification problems, the most important are the following.

Classification of all metrics of homogeneous and Stackel spaces in privileged coordinate systems. For Stackel spaces, this problem was solved in building the theory of complete separation of variables in the papers cited above. For homogeneous spaces, this problem was solved in the work of Petrov (see \cite{27}).

Classification of all (admissible) electromagnetic fields applicable to these methods. For the Hamilton-Jacobi and Klein-Gordon-Fock equations, this problem is completely solved in homogeneous spaces  (see \cite{28}-\cite{33}). In Stackel spaces, it is completely solved for the Hamilton-Jacobi equation and partially solved for the Klein-Gordon-Fock equation  (see \cite{10}-\cite{12}).

Classification of all vacuum and electrovacuum solutions of the Einstein equations with metrics of Stackel and homogeneous spaces in admissible electromagnetic fields. This problem has been completely solved for the Stackel metric (see \cite{13}- \cite{17}). However, this classification problem has not yet been studied for homogeneous spaces.

The solutions to these problems can be viewed as stages of the solution of a single classification problem. In the first two stages, we find all relevant gravitational and electromagnetic fields that are not connected by field equations. In the third stage, using the results of the first two stages, we find metrics and electromagnetic potentials that satisfy the Einstein-Maxwell vacuum equations and have physical meaning.

Thus, for the complete solution to the problem of uniform classification, it remains to integrate the Einstein-Maxwell vacuum equations using the previously found potentials of admissible electromagnetic fields and the known metrics of homogeneous spaces in privileged (canonical) coordinate systems. This problem can also be divided into two stages. In the first stage, all solutions of Maxwell's vacuum equations for the potentials of admissible electromagnetic fields should be found. The present work is devoted to this stage. In the next stage, it is planned to use the obtained results for the integration of the Einstein-Maxwell equations. This will be the subject of further research. The present work is organized as follows.

Section number two contains information from the theory of homogeneous spaces, which will be used later, and definitions and conditions for the potentials of admissible electromagnetic fields, written in canonical frames associated with motion groups of a homogeneous space.

In the third section, Maxwell's vacuum equations are written in canonical frames.

The fourth section contains all solutions of Maxwell's vacuum equations for homogeneous spaces admitting groups of motions \quad $G_3(I)-G_3(VI)$.

\section{Homogeneous spaces}

By the accepted definition, a spacetime manifold \quad $V_4$ \quad is a homogeneous space if a three-parameter group of motions acts on it, whose transitivity hypersurface \quad $V_3$ \quad is endowed with the Euclidean space signature.
Let us introduce a semi-geodesic coordinate system\quad $[u^i] $,\quad in which the metric \quad $V_4$ \quad has the form:
\begin{equation}\label{1}
ds^2 = g_{ij}du^i du^j =-{du^0}^2 + g_{\alpha \beta}du^\alpha du^\beta, \quad det|g_{\alpha\beta}|>0.
\end{equation}
The coordinate indices of the variables of the semi-geodesic coordinate system are denoted by the lower case Latin letters:\quad $i, j, k, l = 0, 1 \dots 3$.\quad The coordinate indices of the variables of the local coordinate system on the hypersurface \quad $ V_3$ \quad are denoted by the lower case Greek letters:\quad $\alpha, \beta, \gamma, \sigma=1, \dots 3.$\quad A 0 index denotes the temporary variable. Group indices and indices of nonholonomic frames are denoted by \quad $a, d, c = 1, \dots 3$.\quad Summation is performed over repeated upper and lower indices within the index range.

There is another (equivalent) definition of a homogeneous space, according to which the spacetime \quad $V_4$ \quad is homogeneous if its subspace \quad $V_3$, \quad endowed with the Euclidean space signature, admits a set of coordinate transformations (the group $G_3$ of motions spaces $V_4$) that allow to connect any two points in \quad $V_3$. \quad (see, e.g., \cite{Landau} ). This definition directly implies that the metric tensor of the $V_3$ space can be represented as follows:
\begin{equation}\label{2}
g_{\alpha\beta}=e^a_\alpha e^b_\beta \eta_{ab},\quad ||\eta_{ab}||=||a_{ab}(u^0)||, \quad e^a_{\alpha, 0}=0,\quad det||a_{ab}||={l_0}^2,
\end{equation}
while the form
$$\omega^a= e^a_\alpha du^\alpha$$
is invariant under the transformation group \quad $G_3$.\quad The vectors of the frame \quad $e^a_\alpha$ \quad (we call them canonical) define a nonholonomic coordinate system in \quad $V_3$,\quad and their dual triplet of vectors
$$
e_a ^\alpha, \quad e_a ^\alpha e^b_\alpha =\delta_a^b, \quad  e_a ^\alpha e^a_\beta = \delta^\alpha_\beta
$$
define the operators of the $G_3$ algebra group:
$$
\hat{Y}_a= e_a ^\alpha \partial_a, \quad [\hat{Y}_a,\hat{Y}_b] = C_{ab}^c \hat{Y}_c.
$$
The Killing vector fields \quad $\xi_a^\alpha $ \quad and their dual vector fields $\xi^a_\alpha$ form another frame in the space $V_3$ (we will call it the Killing frame) and another representation of the algebra of the group \quad $G_3$.\quad In the dual frame, the metric of the space \quad $V_3$ \quad has the form:
\begin{equation}\label{3}
g_{\alpha\beta}=\xi^a_\alpha \xi^b_\beta G_{ab}, \quad \xi_a ^\alpha \xi^b_\alpha =\delta_a^b, \quad  \xi_a ^\alpha \xi^a_\beta = \delta^\alpha_\beta,
\end{equation}
where \quad $G_{ab}$ \quad are the nonholonomic components of the \quad $g_{\alpha\beta}$ \quad tensor in this framework.
The vector fields \quad $\xi_a^\alpha$ \quad satisfy the Killing equations:
\begin{equation}\label{4}
g^{\alpha\beta}_{,\gamma} \xi_a ^\gamma = g^{\alpha \gamma}\xi^\beta_{a,\gamma}+g^{\beta\gamma}\xi^\alpha_{a,\gamma}
\end{equation}
and form the infinitesimal group operators of the algebra \quad $G_3$:
\begin{equation}\label{5}
\hat{X}_a= \xi_a ^\alpha \partial_\alpha, \quad [\hat{X}_a,\hat{X}_b] = C_{ab}^c \hat{X}_c.
\end{equation}
The Killing equation in the \quad $\xi^\alpha_a$ \quad frame has the following form:
\begin{equation}\label{6}
G^{ab}_{|c}=G^{ad}C^b_{dc} + G^{bd}C^a_{dc}  \quad (|a=\xi^\alpha_a \partial_\alpha).
\end{equation}
Indeed, substituting the expression
$$
g^{\alpha\beta}=\xi_a^\alpha \xi_b^\beta G^{ab}
$$
into the equation \eqref{4}, we get
$$
G^{ab}((\xi^\alpha_{a|c}\xi^\beta _b -\xi^\alpha_{a}\xi^\beta_{c|b})+ (\xi^\alpha_{a}\xi^\beta_{b|c}-\xi^\beta_{a}\xi^\alpha_{c|b})) + \xi^\alpha_a\xi^\beta_b G^{ab}_{|c}=0.
$$
Substituting here the commutation relations \eqref{5}, we get:
$$
(G^{ab}_{|c}-G^{ad}C^b_{dc} - G^{bd}C^a_{dc})\xi^\alpha_a\xi^\beta_b=0.
$$

The Hamilton-Jacobi equation for a charged test particle in an external electromagnetic field with potential \quad $A_i$ \quad is:
\begin{equation}\label{7}
  H = g^{ij}P_iP_j=m, \quad P_i=p_i+A_i,\quad p_i=\partial_i\varphi.
\end{equation}
The integrals of motion of the free Hamilton-Jacobi equation are given using Killing vector fields as follows:
\begin{equation}\label{8}
X_a=\xi_a^i p_i,
\end{equation}
Thus, the symmetry of the space given by the Killing vector fields is directly related to the symmetry of the equations of the geodesics given by the integrals of motion. The Hamilton-Jacobi method makes it possible to find these integrals and use them to integrate the geodesic equations. Therefore, the study of the behavior of geodesics is necessary for the study of the geometry of space.

The linear momentum integral of the equation \eqref{7} has the following form:

\begin{equation}\label{9}
  X_a = \xi^i_a P_i + \gamma_a,
\end{equation}
where \quad $\gamma_\alpha$ \quad are some functions of \quad $u^i.\quad $
The equation \eqref{7} admits a motion integral of the form \eqref{8} if \quad $H$ \quad and \quad $\hat{X_a}$ \quad commute under Poisson brackets:
\begin{equation}\label{10}
[H,\hat{X}_a]_P= \frac{\partial H}{\partial p_i}\frac{\partial \hat{X}_a}{\partial x^i} - \frac{\partial H}{\partial x^i}\frac{\partial \hat{X}_a}{\partial p_i}=0 \rightarrow g^{i\sigma}(\xi^{j}_a F_{ji}+\gamma_{a,i})P_\sigma=0.
\end{equation}
Hence:
\begin{equation}\label{11}
\gamma_{a,i} = \xi^{j}_a F_{ij}, \quad F_{ji}=A_{i,j}-A_{j,i}.
\end{equation}
Thus, the admissible electromagnetic field is determined from the equations \eqref{11} (see \cite{31}). In \cite{29}-\cite{30} it was proved that in the case of a homogeneous space, the conditions \eqref{11} can be represented as follows:
\begin{equation}\label{12}
\mathbf{A}_{a|b}= C^c_{ba}\mathbf{A}_{c}, \end{equation}
at the same time
$$
\gamma_a = -\mathbf{A}_{a} \rightarrow \hat{X_a}=\xi^\alpha_a \partial_\alpha.
$$
Here it is denoted:\quad
$$\quad\mathbf{A}_{a}=\xi_a^i A_{i},\quad $$ It can be shown that the equations \eqref{12} form a completely integrable system. This system specifies the necessary and sufficient conditions for the existence of an algebra of integrals of motion that are linear in momenta for equation \eqref{7}. Note that in admissible electromagnetic fields given by the conditions \eqref{12}, the Klein-Gordon-Fock equation
$$
\hat{H}\varphi = (g^{ij}\hat{P}_i\hat{P}_j)\varphi=m^2\varphi, \quad \hat{P}_k=\hat{p}_k+A_k,\quad \hat{p}_k=-\imath\hat{\nabla}_k
$$
also admits an algebra of symmetry operators of the form:
$$
\hat{X}_a = \xi^i_a\hat{\nabla}_i
$$
 (see \cite{29}, \cite{31}).
\quad $\hat{\nabla}_i$ \quad is the covariant derivative operator corresponding to the partial derivative operator -\quad $\hat{\partial}_i =\imath \hat{p}_i$ \quad in the coordinate field \quad $u^i. \quad \varphi$ \quad is a scalar field,  \quad $m=const.$\quad
All admissible electromagnetic fields for the homogeneous spacetime are found in \cite{29}. We will use the results of A.Z. Petrov \cite{27}. We follow the notation used in this book with minor exceptions. For example, the nonignorable variable \quad $x^4$\quad will be denoted \quad $u^0$\quad etc.

\section {Maxwell's equations for an admissible electromagnetic field  in a homogeneous spacetime }

Consider Maxwell's equations with zero electromagnetic field sources in a homogeneous spacetime in the presence of an admissible electromagnetic field:

\begin{equation}\label{13}
\frac{1}{\sqrt{-g}}(\sqrt{-g}F^{ij})_{,j} = 0, \quad g=det|g_{\alpha\beta}|.
\end{equation}
When \quad $i=0$ \quad from the system \eqref{13}, the equation follows:
\begin{equation}\label{14}
\frac{1}{\sqrt{-g}}(\sqrt{-g}g^{\alpha\beta}A_{\beta,0})_{,\alpha} = 0.
\end{equation}
Using the Killing equations \eqref{4}, \eqref{5}, we can obtain:
$$
\frac{g_{|a}}{g}=2\xi^\alpha_{a,\alpha}.
$$
Indeed,
$$
-\frac{g_{|a}}{g}=g^{\alpha\beta}_{|a}g_{\alpha\beta}=G^{bc}_{|a}G_{bc}+2\xi^\alpha_{a,\alpha}+
2C_a =2\xi^\alpha_{a,\alpha}\quad (C_a = C^b_{ab}).$$
Substituting this expression and the relation \eqref{12} into the equation \eqref{14}, we get:
\begin{equation}\label{15a}
G^{ab}C_b\mathbf{A}_{a,0}=0.
\end{equation}
In the case of spaces with groups \quad $G_3(I),\quad G_3(II),\quad G_3(VIII),\quad G_3(IX)\quad C_a=0 $.\quad That is why the equation \eqref{15a} is satisfied.
In the case of the groups \quad $G_3(III),- G_3(VII)$\quad $C_a=const \delta_{a3}$,\quad and from \eqref{15a} it follows:
\begin{equation}\label{15}
 \eta^{3a}\tilde{\mathbf{A}}_{a,0}=0, \quad \tilde{\mathbf{A}}_{a}=A_{\alpha}e^\alpha_a.
\end{equation}
For $i=\alpha$ we have:
\begin{equation}\label{16}
\frac{1}{\sqrt{g}}(\sqrt{g}g^{\alpha\beta}F_{\beta 0})_{,0} +
\frac{1}{\sqrt{g}}(\sqrt{g}g^{\alpha\beta}g^{\gamma\sigma}F_{\beta \sigma})_{,\gamma} = 0.
\end{equation}
We transform the equations \eqref{16} using the \eqref{2} frame. The first term then has the form:
$$
\frac{1}{\sqrt{-g}}(\sqrt{-g}g^{\alpha\beta}F_{\beta 0})_{,0}=-\frac{1}{l_0}(l_0\eta^{ab}\tilde{\mathbf{A}}_{a,0})_{,0}e^\alpha_b, \quad (l_0)^2=det|\eta_{ab}|.
$$
The second term using the \eqref{3} frame, the relations \eqref{12}, and the commutation relations between the operators of the group can be reduced to the following form:
$$
\frac{1}{\sqrt{g}}(\sqrt{g}g^{\alpha\beta}g^{\gamma\sigma}F_{\beta \sigma})_{,\gamma}= \frac{1}{2}G^{a_2 b_1}C^a_{a_2 b_2}(2C_{b_1}G^{bb_2}+C^b_{a_1 b_1}G^{a_1b_2})\xi^\alpha_b\xi^\beta_a e^c_\beta \tilde{\mathbf{A}}_{c}.
$$
So the \eqref{16} equations can be written as follows:
\begin{equation}\label{17}
\frac{1}{l_0}(l_0\eta^{ab}\tilde{\mathbf{A}}_{b,0})_{,0}= \tilde{W}^{ba}\tilde{\mathbf{A}}_{b},
\end{equation}
where
\begin{equation}\label{18}
\tilde{W}^{ab}=(e^a_\beta\xi^\beta_{a_1})(e^a_\alpha \xi^\alpha_{b_1})W^{a_1 b_1}, \quad W^{ab}= \frac{1}{2}G^{a_2 b_1}C^a_{a_2 b_2}(2C_{b_1}G^{bb_2}+C^b_{a_1 b_1}G^{a_1b_2}).
\end{equation}
Then Maxwell's equations can be represented as follows:
\begin{equation}\label{18a}
\beta^a_{,0}= l_0\tilde{W}^{ba}\tilde{\mathbf{A}}_{b},
\end{equation}
\begin{equation}\label{18b}
\tilde{\mathbf{A}}_{a,0} =\frac{1}{l_0}\beta^b \eta_{ab}.
\end{equation}

\section{Maxwell's equations for spaces type I-VI by Bianchi classification}

The group operators in the canonical coordinate set of homogeneous spaces type I- VI by the Bianchi classification can be represented as follows (see [27]):

\begin{equation}\label{19}
X_1=p_1, \quad X_2=p_2, \quad X_3=(ru^1 +\varepsilon u^2)p_1 + nu^2 p_2 -p_3.
\end{equation}
The values \quad $k \quad \varepsilon,\quad n$ \quad for each group take the following values.

\begin{enumerate}
 \item $G(I):\quad k=0, \quad\varepsilon=0,\quad n=0. $

 \item $G(II):\quad k=0,\quad \varepsilon=1,\quad n=0.$

 \item $G(III):\quad k=1,\quad \varepsilon=0,\quad n=0.$

  \item $G(IV):\quad k=1,\quad \varepsilon=1,\quad n=1.$

  \item $G(V):\quad k=1,\quad \varepsilon=0,\quad n=1.$

  \item $G(VI):\quad k=1,\quad\varepsilon=0,\quad n=2.$

  \end{enumerate}
Structural constants can be represented as follows:
\begin{equation}\label{20}
C^c_{ab}=k\delta^c_1(\delta^1_a\delta^3_b -\delta^3_a\delta^1_b)+(\varepsilon\delta^c_1 +n\delta^c_2)(\delta^2_a\delta^3_b -\delta^3_a\delta^2_b)\rightarrow C_a=-(k+n)\delta^3_a
\end{equation}
Find the frame vectors \quad $[\xi^\alpha_a],\quad [e^\alpha_a]$ \quad and their dual vectors \quad $[\xi_\alpha^a],\quad [e_\alpha^a]$. \quad
For this, we use the metrics of homogeneous spaces and the group operators given in \cite{27}.
\begin{equation}\label{21}
\xi^\alpha_a = \delta^1_a \delta^\alpha_1+\delta^2_a \delta^\alpha_2 + \delta^3_a (\delta^\alpha_1(ku^1 + \varepsilon u^2)+ \delta^\alpha_2 nu^2 - \delta^\alpha_3),
\end{equation}
$$
\xi_\alpha^a = \delta_1^a \delta_\alpha^1+\delta_2^a \delta_\alpha^2 + \delta_3^a (\delta_\alpha^1(ku^1 + \varepsilon u^2)+ \delta_\alpha^2 nu^2 - \delta_\alpha^3),
$$

\begin{equation}\label{22}
e^\alpha_a = \delta^1_a \delta^\alpha_1\exp(-ku^3)+\delta^2_a (-\delta^\alpha_1\varepsilon u^3 \exp(-ku^3) + \delta^\alpha_2 \exp(-nu^2)) + \delta^\alpha_3\delta^3_a,
\end{equation}
$$
e^\alpha_a = \delta_1^a \delta_\alpha^1\exp(ku^3)+\delta^2_a (\delta^\alpha_1\varepsilon u^3 \exp nu^3 + \delta^\alpha_2 \exp nu^2)) + \delta_\alpha^3 \delta^3_a.
$$
With these expressions we find the matrix $\tilde{W}^{ab}$ \eqref{18}.
\begin{equation}\label{23}
\tilde{W}^{ab}=\frac{1}{{l_0}^2}[\delta^a_1\delta^b_1(\varepsilon g_{11}+\varepsilon(n-k)g_{12}-kng_{22})\exp(-2nu^3)+
\end{equation}
$$
-(\delta^a_1\varepsilon u^3+\delta^a_2)(\delta^b_1\varepsilon u^3 + \delta^b_2)kn g_{11}\exp(-2ku^3) +
$$
$$
[\delta^b_1(\delta^a_1\varepsilon u^3+\delta^a_2)n(g_{12}+\varepsilon g_{11})) +
\delta^a_1(\delta^b_1\varepsilon u^3 + \delta^b_2)k(g_{12}-\varepsilon g_{11})].
$$
Here (see \cite{27})
$$
g_{11}=a_{11}\exp2ku^3,\quad g_{12}=(\varepsilon u^3 a_{11}+a_{12})\exp(n+k)u^3,\quad g_{22}=(\varepsilon {u^3}^2 a_{11} +2\varepsilon a_{12} +a_{22})\exp2nu^3,\quad
$$
Maxwell's equation \eqref{18a}-\eqref{18b} becomes:
\begin{equation}\label{24}
 \dot{\beta}^b=\frac{1}{l_0}[\delta^a_1\delta^b_1(\varepsilon g_{11}+\varepsilon(n-k)g_{12}-kng_{22})\exp(-2nu^3)+
\end{equation}
$$
-(\delta^a_1\varepsilon u^3+\delta^a_2)(\delta^b_1\varepsilon u^3 + \delta^b_2)kn g_{11}\exp(-2ku^3) +
$$
$$
[\delta^b_1(\delta^a_1\varepsilon u^3+\delta^a_2)n(g_{12}+\varepsilon g_{11})) +
\delta^a_1(\delta^b_1\varepsilon u^3 + \delta^b_2)k(g_{12}-\varepsilon g_{11})]\tilde{\mathbf{A}}_{a}, \quad
$$
\begin{equation}\label{25}
 \beta^a=l_0\eta^{ab}\tilde{\mathbf{A}}_{b,0}.
\end{equation}
The dots denote the time derivatives. The components $\tilde{\mathbf{A}}_{a}$ are defined by the solutions of the \eqref{12} $\mathbf{A}_{b}$ system of equations using the formulas:
\begin{equation}\label{26}
\tilde{\mathbf{A}}_{a}=e^\alpha_a \xi^b_\alpha\mathbf{A}_{b}
\end{equation}
Further solutions of the system of equations \eqref{24} for homogeneous spaces with groups of motions $G_3(I- VI)$ are given. Spatial metrics are given in the book \cite{27}. Solutions for the system \eqref{12} can be found in \cite{28},
$$
\alpha_a=\alpha_a(u^0).
$$

\subsection{Group  $G_3(I)$}

As the parameters \quad $k, n, \varepsilon$ \quad and \quad $C^a_{bc}$ \quad  equal zero, \quad $G_3(I)$ \quad is an Abelian group.
The components of the vector electromagnetic potential  have the form:
$$
\mathbf{A}_{a}=\tilde{\mathbf{A}}_{a}=A_a=\alpha_a,
$$
Substituting these expressions into the system of equations \eqref{24}-\eqref{25}, we obtain the following system of ordinary differential equations:
$$
  \quad \dot{\beta}^a =0 \rightarrow \beta^a= c^a = const ;
$$
$$
l_0\dot{\alpha}_a = a_{ba}c^b\quad \rightarrow \alpha_q =\int\frac{a_{ab}c^{b}}{l_0}du^0,\quad {l_0}^2=det|a_{ab}|.
$$
All components of \quad $a_{ab}$ \quad are arbitrary functions of \quad $u^0$.

\subsection{Group  $G_3(II)$}

For the group $G_3(II)$ the parameters \quad $k, n, \varepsilon$ \quad have the following values: \quad
$
k = n =0, \quad \varepsilon =1.
$
The components of the vector electromagnetic potential in the frames $[\xi^\alpha_a]$ and $[e^\alpha_a]$ have the form:
$$
\mathbf{A}_{1}=\alpha_1, \quad  \mathbf{A}_{2}=\alpha_2 + \alpha_1 u^3, \quad  \mathbf{A}_{3}=\alpha_1 u^3 -\alpha_3; \quad \tilde{\mathbf{A}}_{a}=\alpha_a .
$$
Substituting these expressions into the system of equations \eqref{24}-\eqref{25}, we obtain the following system of ordinary differential equations:
\begin{equation}\label{27}
  \quad   l_0 \dot{\beta}_a = \alpha_1 a_{11}\delta_{1a} \rightarrow l_0 \dot{\beta}_1 = \alpha_1 a_{11}, \quad {\beta}_2=c_2, \quad  {\beta}_3= c_3 \quad (\beta_a=\delta_{ab} \beta^b);
\end{equation}
\begin{equation}\label{28}
l_0\dot{\alpha}_a = a_{1a}\beta_1+a_{2a}c_2 + a_{3a}c_3,\quad {l_0}^2=det|a_{ab}| \quad(c_a=const,).
\end{equation}
The \eqref{27}-\eqref{28} system of equations contains 5 equations for 11 functions:
$$
l_0, \quad a_{ab},\quad \alpha_a, \quad\beta_1.
$$
We should consider separately the variants $\alpha_1=0$ and $\alpha_1\ne 0.$

\quad

1.\quad $\alpha_1=0 \rightarrow \beta_1= c_1 = const.$ \quad In this case, the system of equations \eqref{27} - \eqref{28} has a quadrature solution:
$$
\alpha_q =\int\frac{a_{qb}c_{b_1}\delta^{bb_1}}{l_0}du^0 \quad (q=2,3 ).
$$
For $a=0$, \eqref{28} implies a linear dependence of the components $a_{1q}:$
$$
c_1 a_{11}+c_2 a_{12} + c_3 a_{13} =0.
$$
All independent components of $a_{ab}$ are arbitrary functions of $u^0.$

\quad

2.\quad $\alpha_1\ne 0.$ \quad Consider the following equations from the system \eqref{27} - \eqref{28}:
\begin{equation}\label{29}
l_0\dot{\alpha}_1 =(a_{11}\beta_1+c_2 a_{12}+c_3 a_{13}), \quad l_0\dot{\beta}_1=a_{11}\alpha_1.
\end{equation}
Let us take the function $a_{11}$ out of \eqref{29}. As a result, we obtain:
$$
({\alpha_1}^2 -{\beta_1}^2)_{,0}=
\frac{2\alpha_1}{l_0}(c_2a_{12}+c_3a_{13}).
$$
Hence:
$$
\beta_1 = \xi\sqrt{{\alpha_1}^2 - 2\int\frac{\alpha_1}{l_0}(c_2a_{12}+c_3a_{13})du^0} \quad (\xi^2=1).
$$
From the remaining equations of the system, we get:
$$
\alpha_q = \int\frac{(a_{1q}\beta_1+a_{2q}c_2 + a_{3q}c_3)}{l_0}du^0 \quad (q=2,3);\quad a_{11}=\frac{l_0 \dot{\beta}_1}{\alpha_1}.
$$
The functions \quad $l_{0}$,\quad $\alpha_1$,\quad and all components of \quad $a_{ab}$,\quad except \quad $a_{11},\quad a_{33}$,\quad are arbitrary functions of \quad $u_0$.\quad The component $a_{33}$ results from the equation \quad ${l_0}^2=det|a_{ab}|$:
\begin{equation}\label{30}
a_{33}=\frac{{l_0}^2 +a_{11}{a_{23}}^2 + a_{22}{a_{13}}^2 -2a_{12}a_{13}a_{23}}{a_{11}a_{22}-{a_{12}}^2}
\end{equation}

\subsection{Group  $G_3(III)$}

For the group\quad $G_3(III)$ \quad the parameters \quad $k,\quad n,\quad \varepsilon$ \quad have the following values: \quad
$
k = 1, \quad n =\varepsilon =0.
$ \quad
The components of the vector electromagnetic potential in the frames \quad $[\xi^\alpha_a]$\quad and \quad $[e^\alpha_a]$ \quad have the form:
$$
\mathbf{A}_{1}=\alpha_1\exp u^3, \quad  \mathbf{A}_{2}=\alpha_2, \quad  \mathbf{A}_{3}=\alpha_1 \exp u^3 -\alpha_3.
$$
Substituting these expressions into the system of equations \eqref{24}-\eqref{25}, we obtain the following system of ordinary differential equations:
\begin{equation}\label{31}
  \quad   l_0 \dot{\beta}_a = \alpha_1 a_{12}\delta_{2a} \rightarrow l_0 \dot{\beta}_2 = \alpha_1 a_{12}, \quad {\beta}_1=c_1, \quad  {\beta}_3= 0;
\end{equation}
\begin{equation}\label{32}
l_0\dot{\alpha}_a = a_{2a}\beta_2+a_{1a}c.
\end{equation}
Here and further the equation \eqref{15} is used, according to which \quad $\beta_3=0.$ \quad
The system of equations \eqref{27}-\eqref{28} contains 5 equations for 11 functions:
$$
l_0, \quad a_{ab},\quad \alpha_a, \quad\beta_2.
$$
We should consider separately the variants \quad $\alpha_1=0$ \quad and \quad $\alpha_1\ne 0.$

\quad

1. $\alpha_1=0 \rightarrow \beta_2= c_2 = con                                                                                                                                                  st.$ \quad In this case the system \eqref{27} - \eqref{28} has a solution in quadratures:
$$
\alpha_q =\int\frac{a_{qb}c_{b_1}\delta^{bb_1}}{l_0}du^0 \quad (q=2,3).
$$
From \eqref{28} it follows a linear dependence of the components \quad $a_{1q}:$
$$
c_1 a_{13}+c_2 a_{23}=0 \rightarrow a_{12}=b a_{11}, \quad \beta_1=b, \quad \beta_2=1.
$$
$l_0$ \quad and all independent components of \quad $a_{ab}$ \quad are arbitrary functions of \quad $u^0$. \quad The component \quad $a_{33}$ \quad is found from the equation \eqref{30}

\quad

2. \quad Let \quad $\alpha_1\ne 0.$ \quad Consider the following equations from the system \eqref{27} - \eqref{28}:
\begin{equation}\label{33}
l_0\dot{\alpha}_1 =a_{12}\beta_2+c_1 a_{11}, \quad l_0\dot{\beta}_2=a_{12}\alpha_1.
\end{equation}
from the system \eqref{33} it follows:
$$
({\alpha_1}^2 -{\beta_2}^2)_{,0}= \frac{2\alpha_1}{l_0}c_1a_{11}.
$$
Hence:
$$
\beta_2 = \xi\sqrt{{\alpha_1}^2 - 2\int\frac{\alpha_1}{l_0}(c_1a_{11}+c_3 a_{13})du^0} \quad (\xi^2=1).
$$
From the remaining equations of the system we get:
$$
\alpha_q = \int\frac{(a_{2q}\beta_2+a_{1q}c_1 + a_{3q}c_3)}{l_0}du^0 \quad (q=2,3);\quad a_{11}=\frac{l_0 \dot{\beta}_2}{\alpha_1}.
$$
The functions \quad $l_{0}$,\quad $\alpha_1$\quad and all components of \quad $a_{ab}$,\quad except \quad $a_{11},\quad a_{33}$,\quad are arbitrary functions of \quad $u_0$.\quad The component \quad $a_{33}$ \quad results from the equation \eqref{30}

\subsection{Group $G_3(IV)$}

For the group \quad $G_3(IV)$ \quad the parameters \quad $k,\quad n,\quad \varepsilon$ \quad have the values: \quad
$
k = n =\varepsilon =1.
$ \quad
The components of the vector electromagnetic potential in the frames \quad$[\xi^\alpha_a]$\quad and \quad $[e^\alpha_a]$ \quad have the form:
$$
\mathbf{A}_{1}=\alpha_1\exp u^3, \quad  \mathbf{A}_{2}=(\alpha_2+\alpha_1u^3)\exp u^3, \quad  \mathbf{A}_{3}=(\alpha_1(u^1 +u^2+u^2u^3)+\alpha_2u^2)\exp u^3 -\alpha_3; $$
$$
\tilde{\mathbf{A}}_{a}=\alpha_a.
$$
Maxwell's equations \eqref{18a}, \eqref{18b} reduce to the following system:
\begin{equation}\label{34}
l_0\dot{\beta}_a=\delta_{1a}(a_{11}(\alpha_1 +\alpha_2)-\alpha_1a_{22}+\alpha_2a_{12})+\delta_{2a}(\alpha_1a_{12}
-a_{11}(\alpha_1 +\alpha_2)).
\end{equation}
\begin{equation}\label{35}
l_0\dot{\alpha}_a=\beta_2a_{a2}+\beta_1a_{a1}, \quad \beta_3=0.
\end{equation}
from the system \eqref{35} it follows:
\begin{equation}\label{36}
\dot{\alpha}_3=\int\frac{\beta_2a_{32}+\beta_1a_{31}}{l_0}du^0.
\end{equation}
Let us now consider the remaining equations.

\quad

A)\quad  $\beta_1 \ne 0.$

From the system \eqref{34} it follows:
\begin{equation}\label{38}
a_{12}=\frac{1}{\beta_1}(l_0\dot{\alpha}_2-\beta_2 a_{22}) \quad a_{11}=\frac{1}{{\beta_1}^2}(l_0(\dot{\alpha}_1\beta_1-\dot{\alpha}_2\beta_2)+     {\beta_2}^2 a_{22}),
\end{equation}
\quad
Using these relations, we obtain a consequence from the remaining equations of the system \eqref{34}-\eqref{35}:
\begin{equation}\label{39}
\beta_1\dot{\beta}_2-\beta_2(\dot{\beta}_1 + \dot{\beta}_2)=\alpha_1\dot{\alpha}_2-(\alpha_1+\alpha_2)\dot{\alpha}_1.
\end{equation}
With the equation \eqref{39} the dependent functions \quad $\alpha_a,\quad \beta_a$\quad can be expressed in terms of the independent functions. Let us write down the solutions.

\quad

1.\quad $(\alpha_1\beta_1 + \beta_2(\alpha_1+\alpha_2))\beta_2 \ne 0.$
$$
\beta_1= \beta_2(b-\ln{\beta_2} - \int{\frac{\alpha_1\dot{\alpha}_2-(\alpha_1+\alpha_2)\dot{\alpha}_1}{{\beta_2}^2}}du^0);
$$
$$
a_{22}=\frac{l_0(\dot{\alpha}_2(\alpha_1+\alpha_2)-\beta_1(\dot{\beta}_1 + \dot{\beta}_2))}{\alpha_1\beta_1 + \beta_2(\alpha_1+\alpha_2)}.
$$
$l_0,\quad a_{13},\quad a_{23},\quad \varphi$\quad are arbitrary functions of time. \quad The function \quad $a_{33}$ \quad is expressed in terms of these functions using the relation \eqref{30}

\quad

2. \quad $\alpha_1\beta_1 + \beta_2(\alpha_1+\alpha_2) = 0, \quad a_{22}$,\quad is an arbitrary function, depending on \quad $u^0$.
$$
\alpha_1=a\exp{\varphi}+b\exp{\varphi}, \quad \alpha_2=(1+e)\alpha_1 \quad \beta_2=a\exp{\varphi}-b\exp{\varphi},\quad \beta_1=e\beta_2\quad (e=const).
$$
$l_0,\quad a_{13},\quad a_{23},\quad \varphi$\quad are arbitrary functions of time. \quad The function \quad $a_{33}$ \quad is expressed in terms of these functions using the relation \eqref{30}.

\quad

3.\quad $\beta_2=0$.
$$
\alpha_2=\alpha_1(a + \ln{\alpha_1}), \quad a_{12}=\frac{l_0\dot{\alpha}_2}{\beta_1}, \quad a_{11}=\frac{l_0\dot{\alpha}_1}{\beta_1},
\quad a_{22}=\frac{l_0(\dot{\alpha}_2(\alpha_1+\alpha_2)-\dot{\beta}_1\beta_1)}{\alpha_1\beta_1}
$$
$l_0,\quad a_{13},\quad a_{23},\quad \alpha_1, \quad \beta_1 $\quad are arbitrary functions of time. \quad The function \quad $a_{33}$ \quad is expressed in terms of these functions using the relation \eqref{30}.

\quad

B) \quad $\beta_1=0.$ \quad Maxwell's equations take the form:

\quad

$$
l_0\dot{\beta}_2 = \alpha_1 a_{12}-(\alpha_1+\alpha_2)a_{11},\quad
l_0\dot{\beta}_2 = -\alpha_1 a_{22}+(\alpha_1+\alpha_2)a_{12};
$$
$$
l_0\dot{\alpha}_1=\beta_2a_{12},\quad   l_0\dot{\alpha}_2=\beta_2a_{22}.
$$
The system has the following solution:

\quad

a) \quad $(\alpha_1+\alpha_2)\ne0$.

$$
\beta_2=\xi\sqrt{b + 2\int{\frac{1}{l_0}(\dot{\alpha}_1(\alpha_1 + \alpha_2)-\alpha_1\dot{\alpha}_2)}du_0}. \quad a_{12}=\frac{l_0\dot{\alpha}_1}{\beta_2}, \quad a_{22}=\frac{l_0\dot{\alpha}_2}{\beta_2}.
$$

$$a_{11}=\frac{l_0(\alpha_1\dot{\alpha}_1-
\beta_2\dot{\beta}_2)}{\beta_2(\alpha_1+\alpha_2)}$$
$l_0, \quad a_{13},\quad a_{23}, \quad \alpha_1, \quad \alpha_2$ \quad are arbitrary functions of time. The function \quad $a_{33}$ \quad is expressed in terms of these functions using the relation \eqref{30}.

\quad

b) \quad $\alpha_2 = -\alpha_1 \rightarrow \alpha_1=a\exp{\varphi}-b\exp{\varphi}\quad \beta_2=a\exp{\varphi}+b\exp{\varphi},\quad
a_{12}=\frac{l_0\dot{\alpha}_1}{\beta_2}\quad a_{22}=\frac{l_0\dot{\alpha}_2}{\beta_2}.$

$l_0,\quad a_{11},\quad a_{13},\quad a_{23},\quad\varphi,\quad \beta_1 $\quad are arbitrary functions of time. The function \quad $a_{33}$ \quad is expressed in terms of these functions using the relation \eqref{30}.

\subsection{Group $G_3(V)$}

For the group \quad $G_3(V)$ \quad the parameters \quad $k,\quad n,,\quad \varepsilon$ \quad have the values:\quad $k = n =1, \quad \varepsilon =0.$
The components of the vector electromagnetic potential in the frames \quad $[\xi^\alpha_a]$\quad and \quad $[e^\alpha_a]$ \quad have the form:
$$
\mathbf{A}_{1}=\alpha_1\exp u^3, \quad  \mathbf{A}_{2}=\alpha_2\exp u^3, \quad  \mathbf{A}_{3}=(\alpha_1 u^1 + \alpha_2 u^2)\exp u^3 -\alpha_3; $$
$$
\tilde{\mathbf{A}}_{a}=\alpha_a.
$$
Maxwell's equations \eqref{17} reduce to the following system of equations:
\begin{equation}\label{40}
l_0\dot{\alpha}_a=\beta_2a_{a2}+\beta_1a_{a1}, \quad \beta_3=0.
\end{equation}
\begin{equation}\label{41}
l_0\dot{\beta}_a=\delta_{1a}(a_{12}\alpha_{2} -\alpha_1 a_{22})+ \delta_{2a}(a_{12}\alpha_{1} - a_{11}\alpha_2),
  \end{equation}

Hence:
$$
\dot{\alpha}_3=\int\frac{\beta_2a_{32}+\beta_1a_{31}}{l_0}du^0,
$$
\begin{equation}\label{42}
l_0\dot{\alpha}_1=(a_{11}\beta_1+a_{12}\beta_2), \quad l_0\dot{\alpha}_2=(a_{12}\beta_1+a_{22}\beta_2).
\end{equation}

 1.\quad $\alpha_1 \ne 0$. \quad From the system \eqref{41} it follows:
\begin{equation}\label{43}
a_{12}=\frac{1}{\alpha_1}(l_0\dot{\beta}_2+\alpha_2 a_{11}), \quad a_{22}=\frac{1}{{\alpha_1}^2}(l_0(\dot{\beta}_2\alpha_2 -\dot{\beta}_1\alpha_1) +a_{11}{\alpha_2}^2).
\end{equation}
Substituting \eqref{43} into \eqref{42}
, we get the corollary:
\begin{equation}\label{44}
\beta_1\dot{\beta}_2-\beta_2\dot{\beta}_1 =\alpha_1\dot{\alpha}_2-\alpha_2\dot{\alpha}_1.
\end{equation}
\begin{equation}\label{45}
a_{11}(\alpha_1\beta_1 + \alpha_2\beta_2)=l_0(\dot{\alpha}_1\alpha_1 - \dot{\beta_2}\beta_2).
\end{equation}

From \eqref{44} it follows:
$$
\alpha_2=\alpha_1(b+\int{\frac{\beta_1\dot{\beta}_2-\beta_2\dot{\beta}_1}{{\alpha_1}^2}}du^0), $$
Let us consider \eqref{45}.

\quad

a)\quad $\alpha_1\beta_1 + \alpha_2\beta_2 \ne 0.$ \quad Then we have:

$$
a_{11}=\frac{l_0(\alpha_1\dot{\alpha}_2-\alpha_2\dot{\alpha}_1)}{\alpha_1\beta_1 + \alpha_2\beta_2};
$$
$l_0, \quad a_{13},\quad a_{23}, \quad \alpha_1, \quad\beta_a$ \quad are arbitrary functions of time. The function \quad $a_{33}$ \quad is expressed in terms of these functions using the relation \eqref{30}.

\quad

b)\quad $ \alpha_1\beta_1 + \alpha_2\beta_2 = 0\quad \rightarrow \alpha_1\dot{\alpha}_1 - \beta_1\dot{\beta}_1=0,\quad
\alpha_1\dot{\alpha}_2 + \beta_2\dot{\beta}_1=0.$

\quad

From this, it follows:
$$
\alpha_1=a\exp{\varphi}+b\exp{\varphi}, \quad \beta_2=a\exp{\varphi}-b\exp{\varphi}, \quad \alpha_2=-l\alpha_1, \quad \beta_1=l\beta_2,
$$
where \quad $a,\quad b,\quad l =const, \quad \varphi=\varphi(u^0).\quad$
$l_0,\quad a_{11}, \quad a_{13},\quad a_{23}$ \quad are arbitrary functions of time. The function \quad $a_{33}$ \quad is expressed in terms of these functions using the relation \eqref{30}.

\quad

2.\quad  $\alpha_1 = 0$. \quad From the system \eqref{41} it follows:
$$
a_{12}=\frac{l_0 \dot{\beta}_1}{\alpha_2}, \quad a_{11}=-\frac{l_0 \dot{\beta}_2}{\alpha_2}, \quad
a_{22}=\frac{l_0 (\dot{\alpha}_2\alpha_2-\dot{\beta}_1\beta_1)}{\alpha_2\beta_2},\quad \beta_1=a\beta_2,
$$
here \quad
$a= const, \quad l_0,\quad  a_{13},\quad a_{23}, \quad \alpha_2,\quad \beta_2$ \quad are arbitrary functions of time. The function \quad $a_{33}$ \quad is expressed in terms of these functions using the relation \eqref{30}.

\subsection{Group $G_3(VI)$}

For the group \quad $G_3(VI)$ \quad the parameters \quad $k,\quad n,\quad \varepsilon$ \quad have the following values: \quad
$
k =1 \quad n =2, \quad \varepsilon =0.
$ \quad
The components of the vector electromagnetic potential in the frames \quad $[\xi^\alpha_a]$ \quad and \quad $[e^\alpha_a]$ \quad have the form:
$$
\mathbf{A}_{1}=\alpha_1\exp u^3, \quad  \mathbf{A}_{2}=\alpha_2\exp 2u^3, \quad  \mathbf{A}_{3}=\alpha_1 u^1\exp u^3 + 2\alpha_2 u^2\exp2u^3 -\alpha_3; $$
$$
\tilde{\mathbf{A}}_{a}=\alpha_a.
$$
Maxwell's equation \eqref{17} has the form:
\begin{equation}\label{45}
l_0\dot{\alpha}_a=\beta_2a_{a2}+\beta_1a_{a1}.
\end{equation}
\begin{equation}\label{46}
l_0\dot{\beta}_a=\delta_{1a}(a_{12}\alpha_{2} -2\alpha_1 a_{22})+ \delta_{2a}(a_{12}\alpha_{1} - 2a_{11}\alpha_2),\quad \beta_3=0,
\end{equation}

from the system \eqref{45} it follows:
$$
\dot{\alpha}_3=\int\frac{\beta_2a_{32}+\beta_1a_{31}}{l_0}du^0.
$$
\begin{equation}\label{47}
l_0\dot{\alpha}_1=(a_{11}\beta_1+a_{12}\beta_2), \quad l_0\dot{\alpha}_2=(a_{12}\beta_1+a_{22}\beta_2).
\end{equation}
 I. \quad $\beta_1 \ne 0 $, \quad from the system \eqref{45} it follows:
\begin{equation}\label{48}
a_{12}=\frac{1}{\beta_1}(l_0\dot{\alpha}_2-\beta_2 a_{22}), \quad a_{11}=\frac{1}{{\beta_1}^2}(l_0(\dot{\alpha}_1\beta_1 -\dot{\alpha}_2\beta_2) +a_{22}{\beta_2}^2).
\end{equation}
Substituting  \eqref{48} into \eqref{45}, we get:
\begin{equation}\label{49}
a_{22}(\alpha_1\beta_1 + 2\alpha_2\beta_2)=l_0(\alpha_2\dot{\alpha}_2 - \dot{\beta}_1\beta_1),\quad
\end{equation}

\begin{equation}\label{50}
(2\alpha_1\beta_ 1 + \alpha_2\beta_2)(2\alpha_2\dot{\alpha}_1 + \dot{\beta}_2\beta_1)=(\dot{\beta}_1\beta_2 + 2\dot{\alpha}_2\alpha_1)(\alpha_1\beta_1 + 2\alpha_2\beta_2)=0
\end{equation}
Using this relation, we get the following solutions:

\quad

1) \quad $\alpha_1\beta_1 + 2\alpha_2\beta_2 \ne 0. \quad $ From \eqref{49} it follows: \quad $$
a_{22}=\frac{l_0(\dot{\alpha}_2\alpha_2-\dot{\beta}_1\beta_1)}{(\alpha_1\beta_1 + 2\alpha_2\beta_2)}.
$$
Denote:
$$
\alpha_q=a_q\exp{\varphi}. \quad \beta_q=b_q\exp{\varphi} \quad (q=1,2),
$$
where \quad $ a_q,\quad b_q,\quad \varphi $ \quad are  functions of \quad $u^ 0.$ \quad From the equation \eqref{50} we get:
$$
\dot{\varphi}=\frac{(\dot{b}_1b_2 + 2\dot{a}_2 a_1)(a_1 b_1 + 2a_2b_2)-(2a_1 b_ 1 + a_2b_2)(2a_2\dot{a}_1 + \dot{b}_2 b_1)}{(2a_1 a_2 +b_1 b_2)(a_1b_1-a_2b_2)};
$$

$$
a_{12}=\frac{l_0(\dot{\varphi} a_2 +a_2)-b_2 a_{22}}{b_1};
\quad  a_{11}=\frac{l_0((a_1b_1-a_2b_2)\dot{\varphi}+\dot{a}_1b_1-  \dot{a}_2b_2) +{b_2}^2 a_{22}}{{b_1}^2};
$$
$$
a_{22}=\frac{l_0((a_2^2-b_1^2)\dot{\varphi}+\dot{a}_2a_2-\dot{b}_1b_1)}{2a_1b_1+a_2b_2}.
$$

$l_0,\quad a_{13},\quad a_{23},\quad a_q,\quad b_q$ \quad are arbitrary functions dependent on time. The function $a_{33}$ is expressed by these functions using the relation \eqref{30}

\quad

2)
$\quad \dot{\alpha}_2\alpha_2-\dot{\beta}_1\beta_1=0 \quad\rightarrow \quad \alpha_1\beta_1 + 2\alpha_2\beta_2=0.\quad a_{22}$ - is an arbitrary function from $u^0;$
$$
\alpha_2=a\exp\varphi -b\exp(-\varphi), \quad \beta_1=a\exp\varphi +b\exp(-\varphi).
$$
From this, it follows:

\quad

a)

$$
\quad \alpha_1 =-\frac{\beta_2}{2}(\frac{a\exp\varphi -b\exp(-\varphi)}{a\exp\varphi+b\exp(-\varphi)});\quad
$$
$$
a_{12}=l_0\dot{\varphi}-\frac{\beta_2 a_{22}}{\beta_1},\quad a_{11}=\frac{l_0(\dot{\alpha}_1\beta_1-\dot{\alpha}_2\beta_2)+\beta_2^2a_{22}}{\beta_1^2}
$$

\quad

b) \quad $\dot{\varphi}=0 $
$$\beta_1=1,\quad \alpha_2=-2b,\quad \alpha_1=-b\beta_2, \quad
a_{12}=-\beta_2 a_{22}, \quad a_{11}=-bl_0\dot{\beta}_2 +{\beta_2}^2a_{22}.
$$
where $\quad l_0,\quad a. \quad b =const \quad a_{22},\quad a_{13},\quad a_{23}, \quad \beta_2,\quad \varphi $ \quad are arbitrary functions dependent on time.

\quad

II.\quad $\beta_1=0$.\quad
From \eqref{45}-\eqref{46} it follows:
\begin{equation}\label{51}
a_{12}=\frac{2l_0\dot{\alpha}_2 \alpha_2}{\beta_2},\quad
a_{22}=\frac{l_0\dot{\alpha}_2}{\beta_2},\quad   a_{11}=\frac{l_0(2b^2\dot{\alpha}_2{\alpha_2}^3-\beta_2\dot{\beta}_2)}{2\alpha_2\beta_2},\quad
\quad \alpha_1=b{\alpha_2}^2.
\end{equation}

$l_0,\quad a_{22},\quad a_{13},\quad a_{23}, \quad \alpha_2 \quad \beta_2$ \quad depends arbitrarily on time functions. The function $a_{33}$ is expressed in terms of these functions using the relation \eqref{30}.

\section{Conclusion}

The performed classification of admissible electromagnetic fields will be used in the search for electrovacuum solutions of the Einstein-Maxwell equations. As known, the components of the Ricci tensor of a homogeneous space in the frame \eqref{2} depend only on time. In order for Einstein's equations with matter to prove to be an integrable system of ordinary differential equations, the equations of motion of matter must be subordinated to the conditions of space symmetry. These conditions are fulfilled first by the potentials of the electromagnetic fields determined in this work.

\end{document}